# Experimental observation of drift wave turbulence in an inhomogeneous six-pole cusp magnetic field of MPD

A. D. Patel[*], M. Sharma, R. Ganesh, N. Ramasubramanian, and P. K. Chattopadhyay

*Institute for Plasma Research, HBNI, Bhat, Gandhinagar, Gujarat - 382428, India*

*E-mail: *amitpatel@ipr.res.in*

**Abstract:**

This paper presents a detailed study on the controlled experimental observation of drift wave instabilities in an inhomogeneous Six pole cusp magnetic field generated by an in-house developed Multi-pole line cusp magnetic field device (MPD) [Patel *et al*. Rev. Sci. Instrum., **44**, 726 (2018)]. The device is composed of six axially symmetric cusps and non-cusp (in between two consecutive magnets) regions. The observed instability has been investigated in one of these non-cusp regions by controlling the radial plasma density gradient with changing pole magnetic field which is a unique feature of this device. It has been observed that the frequency of the instability changes explicitly with the density gradient. Moreover the scale length of plasma parameters, frequency spectrum, cross-correlation function, and fluctuation level of plasma densities has been measured in order to identify the instability. The cross field drift velocity due to fluctuation in plasma parameters have been measured from the wave number- frequency $S(k_z, \omega)$ spectrum and verified with the theoretical values obtained from density scale length formula. Further from the $S(k_z, \omega)$ spectrum it has been found that the drift velocity alternates the sign in the consecutive non-cusp regions.

## I. Introduction

Confinement of plasma in a multi pole cusp magnetic field has been widely used in laboratory devices for basic plasma studies because of their ability to confine a large amount of uniform quiescent plasma [1, 2]. Although the studies of plasma confinement using multi-cusp configuration originated from fusion research [3- 4], later it has been utilized as a useful paradigm in order to study different plasma properties in different devices. Ion beam sources [5], plasma-etching reactors [6] and plasma wave studies [1] are examples of such kind of devices. Moreover, it has been observed that the diffusion across the inhomogeneous cusp magnetic field explicitly controls the efficiency of the plasma devices [3, 7-8]. Thus, the study of plasma diffusion across an inhomogeneous magnetic field has received a great deal of attention in both laboratory and space plasma experiments. This diffusion across the inhomogeneous magnetic field in fusion devices (tokamaks), reverse field pinches, stellarators, linear plasma devices [9-11] and the cusp magnetic field [12-14] devices produce a large number of instability mechanisms, which results in anomalous transport of plasma particles across the field. Drift wave turbulence is thought to be the dominant source of anomalous transport in the aforementioned devices [9-12, 15-16]. Thus drift wave instabilities are considered as "universal instabilities" generated in magnetically confined plasma devices. Drift wave instabilities are the dominant mechanism for the transport of particles, energy and momentum across magnetic field lines. These instabilities have relatively longer wavelengths and thus these instabilities are the key reasons for "anomalous diffusion" across the inhomogeneous magnetic field. Here we note that this drift is the diamagnetic drifts associated with the density (or temperature) gradients.

Now in order to understand the physics of plasma diffusion across an inhomogeneous magnetic field, several controlled laboratory experiments have been carried out in last four decades [3, 17]. There exist a number of ways to produce an inhomogeneous magnetic field either using permanent magnets or electro magnets [3, 17-20]. The most of earlier control laboratory studied on drift wave turbulence has been carried out in linear devices [21-25]. The controlled experiments [22-24] (in homogeneous magnetic field) have been performed in low temperature, low–β (*is ratio of plasma pressure to magnetic pressure*) $\ll \sqrt{m_e/m_i}$ ), in fully ionized Cs and K plasmas. In these devices drift wave oscillations in a radial region of the plasma column coinciding with a density gradient have been reported and it has been found that these waves are density gradient driven drift waves. In 1974, D'Angelo *et al.* [25] , studied the drift wave turbulence in Cs plasma confined in multi cusp magnetic field and reported the spectral properties of density fluctuation of drift wave turbulence and also notice that drift velocity is inversely proportional to magnetic field. Latter in 1986, Bosh *et al.* [12], studied the drift wave turbulence in cusp region of spindle cusp magnetic field and reported that the plasma diffusion across the magnetic field in the cusp occur by combined effect of neutral-charged particle collisions and Bohm diffusion. Here we would like to mention that the experimental observation of drift wave turbulence across the inhomogeneous magnetic field of complex geometry (cusp magnetic field) is uncovered till date [3, 12, 25]. Most of the devices reported before with multi cusp configurations had very small volume of low beta region along the edge so as to maximize the usage of the central high beta region. Hence the various fluctuations present in the edge regions (low-$\beta$) and their effects on the central region (high-$\beta$) have not been studied before.

In our experiment, we have developed an experimental **M**ulti-pole line cusp **P**lasma **D**evice (MPD) [26], in which multi line cusp magnetic field have been produced by placing six electromagnets and magnetic field lines are again profiled using Vacoflux-50 core material. The device also constructed in such a configuration that the volume of edge region (low-$\beta$) is comparable to the volume of central region (high-$\beta$). The argon plasma is produced by hot cathode filament based source located at the centre of one end of the chamber. The confined (bulk) plasma at the centre is high dense, hot and enrich of primary electrons (because cusp magnetic field confine primary electron in confined region). This bulk plasma diffuses across the magnetic field and stream out along magnetic field. Because of diffusion we have observed gradient in plasma density across the magnetic field along the non-cusp region and at the edge of the device. The bulk plasma at the central region of device is quiescent ($\delta n/n < 0.5$ %) and radially outward from the central region across magnetic field the gradient in density increases, thus the fluctuation in plasma density also increase accordingly. In addition the primary non thermal electrons (which act as an additional source of free energy to driving up fluctuations [27-28]) are scavenged across the magnetic field and cannot diffuse to our studying plasma region.

A brief description of the device along with the diagnostics has been given in section 2. Then in section 3 we explicitly study the drift wave instabilities due to density gradient produced in an inhomogeneous magnetic field of non-cusp region [26]. At the outset, we present the radial variation of the mean plasma parameters. From the radial variation of plasma parameter we compare inequality of scale length of plasma parameter to estimate the instability. Next, we compare the time profile of plasma density fluctuations due to density gradient with time profile of plasma density fluctuation of quiescent bulk plasma confined in central region of the device. The measured radial profile of normalized density fluctuation level follows the radial density gradient as expected for drift wave. Then using this time profile of density fluctuation we discuss the frequency spectra of the instability. We also present the cross correlation function between the time profiles of normalized density and floating potential fluctuation. Further, the cross field drift velocity due to fluctuations in plasma parameters have been measured from the wave number- frequency ($S(k_z, \omega)$) spectrum and verified with the theoretical values obtained from density scale length formula. From the $S(k_z, \omega)$ spectrum of each of non cusp region it has been found that the drift velocity alternates the sign in the consecutive non-cusp regions. Finally in section 4 we summarised and conclude this article.

## II. Experimental Setup and plasma diagnostic:

a) **Experimental Setup**

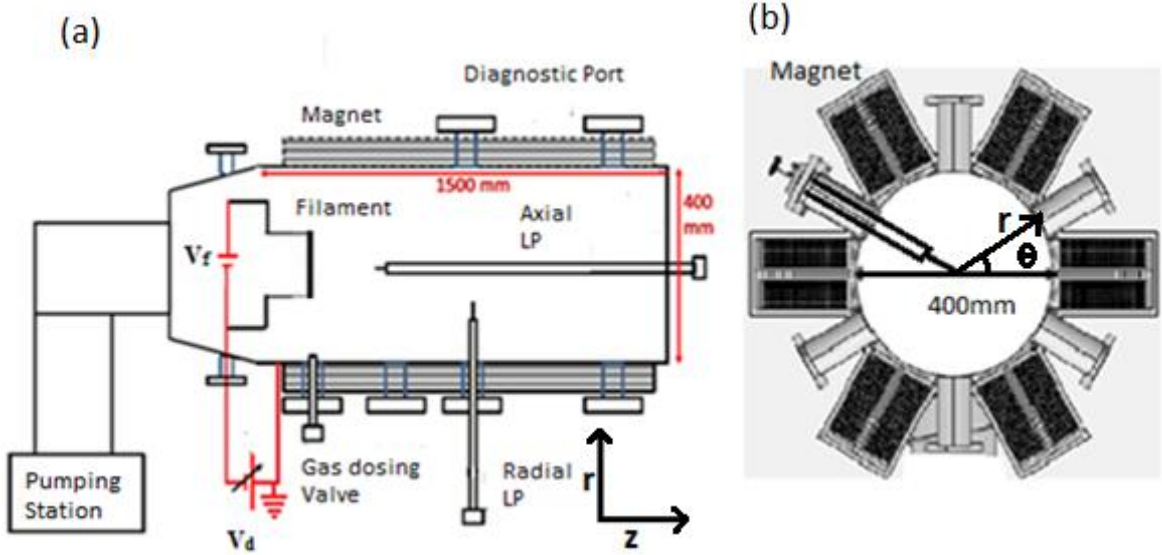

Figure 1. Schematic diagram of (a) Experimental setup and (b) Chamber end cross sectional view of the multi-line cusp magnetic field plasma device, where LP-Langmuir Probe, $V_f$- floating Power supply for filament heating, $V_d$- discharge power supply.

Figure 1, (a) shows a Schematic diagram of experimental setup and (b) shows end cross sectional view of experimental set up of the variable multi cusp magnetic field plasma device (MPD). The experimental setup consists of vacuum chamber (diameter = 40 cm and length = 1.5 m) with wall thickness 0.6 cm. This chamber is pumped out by a combination of Rotary-TMP (430 liter/Sec) pump capable of $10^{-6}$ mBar of base pressure. The cusp magnetic field has been produced by six rectangular electromagnets and magnetic field is profile using vacoflux-50 core material [7]. Figure 2 shows the 2-D contour plot of the magnetic field, simulated using Finite Element Method Magnetics (FEMM) [29] on the centre $(r, \theta)$ plane of the chamber in vacuum condition. The radial variation of the measured magnetic field $(B_M)$ along the cusp region (θ =0°) and the non-cusp region (θ = 30°) are shown in figure 3 for two different currents ($I_{magnet}$ = 100 A, and 150 A) passes through the electromagnets. Figure 3 shows the comparison of magnetic field between measured $(B_M)$, simulated $(B_S)$ shows good match. The filamentary Argon discharge plasma was produced using hot filament based cathode source. The plasma source (cathode) is two dimensional (8cm x 8cm) vertical array of five tungsten filaments; each filament has 0.5 mm diameter and 8 cm length. It is fitted from the conical reducer such that the filaments are inside the main chamber itself where the magnetic field is low. Also it has been taken care to push the source well inside the main chamber to avoid the edge effects of the magnets. These filaments are powered by a 500 A, 15 V floating power supply ($V_f$) while it is normally operated at around 16 - 19 A per filament. The chamber was filled with Argon gas through a needle valve to a pressure 2 x $10^{-4}$ mBar. The source is biased with a voltage of - 76 V with respect to the grounded chamber walls using discharge power supply ($V_d$). The primary electrons emitted from the filaments travel in the electrical field directions, while they are confined by the cusp magnetic field lines. Because of mirror effects due to cusp configuration, electrons move back and forth between the poles ionized the background gas atoms. The confinement of plasma is provided by a multi-cusp magnetic field produced by a six electromagnets accommodated on chamber surface. It also enhances the confinement of primary electrons and subsequently helps in density enhancement.

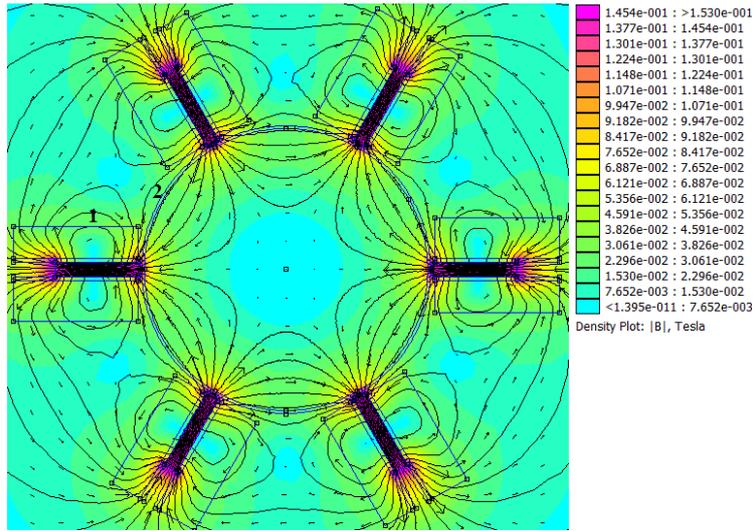

Figure 2. Contour plot of the vacuum field lines on $(r, \theta)$ plane of device using FEMM simulation when magnets are energized with 150A current. 1 is electromagnet, 2 is chamber cross section.

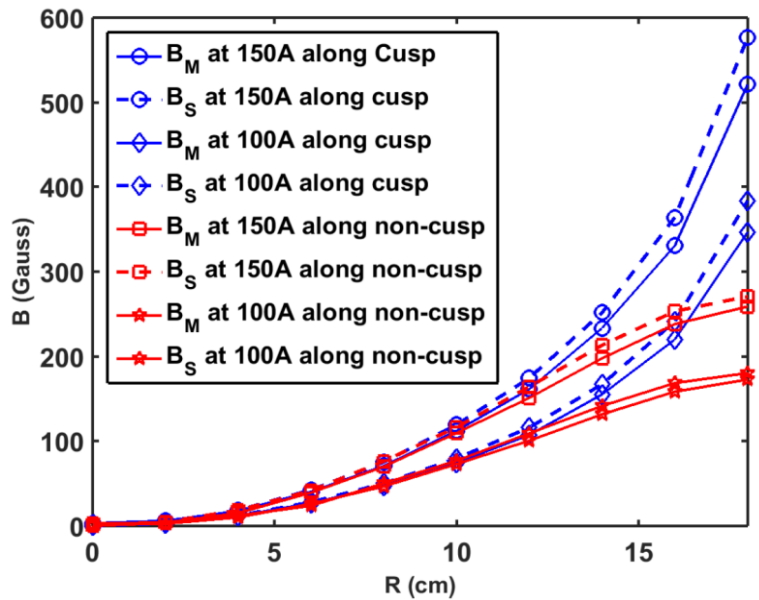

Figure 3. Radial variation of magnetic field (both simulated ($B_s$) and measured ($B_m$)) along the cusp and non-cusp regions when magnets are energized with two different currents ($I_{mag}$) 100 A and 150 A.

b) **Plasma Diagnostic**

For measuring the radial variation of mean value of plasma density ($n$), electron temperature ($T_e$), and floating potential ($V_f$) from *I-V* characteristic of plasma obtained using simple Langmuir probe a sweeping circuit of frequency 1 Hz and a voltage sweep from -80 V to +30 V has been used. An analysis code has been developed in MATLAB for quick processing of the data. For calculation of mean electron temperature ($T_e$) and plasma density ($n$) following equations is used throughout the manuscript.

In general, bi-Maxwellian plasmas consist of two types of electrons with hot electron temperature component ($T_h$) and cold electron temperature component ($T_{ec}$) [30-31]. Kinetic definition of Effective temperature ($T_e$) can be expressed as:

$$T_e = \frac{1}{3} m_e \int_{-\infty}^{\infty} v^2 f(v) dv \dots \dots \dots \dots (1).$$

An effective temperature can be derived by assuming electron energy distribution function (EEDF) as bi-Maxwellian, $f(v) = \alpha f_h(v) + (1-\alpha) f_c(v)$, Where $f_h(v)$ is the hot electron distribution function and $f_c(v)$ is the cold electron distribution function. $\alpha$ is the ratio of hot electron current to electron current at plasma potential [29-30]. Hence, for bi-Maxwellian plasma, the effective temperature can be derived as [30]:

$$T_e = (1-\alpha)T_{ec} + \alpha T_h \quad\ldots\ldots\ldots\ldots\ldots\ldots\ldots\ldots\quad (2).$$

A detailed analysis of electron temperature ($T_e$) has been reported in an earlier publication by the same authors [26].

In our experiments, the Debye number (ratio of the probe radius to the Debye length) varies from 3 to 10. The plasma density ($n$) is determined from the ion saturation current by assuming the plasma to be quasi-neutral ($n_e \approx n_i = n$). The details of density ($n$) determination are given in Ref. 23 and 32.

## III. Experimental results and discussion:

### a) Equilibrium Profiles

The device has facilities to change the multi cusp magnetic field values by changing the magnet current and thus associated equilibrium profiles of plasma parameters are also changed. Out of various configurations of equilibrium profiles we have limited our measurements to two extreme cases for studying mean plasma parameters and their fluctuation. In this present study we have changed the multi cusp magnetic field values at two different magnet currents 100 A, and 150 A. The corresponding magnetic field value at the pole of cusp (near the magnet, R = 20 cm) is 0.75 kG and 1.1 kG respectively. The radial variation of plasma parameters and its fluctuation (plasma densities, electron temperature, and floating potential) for these two field values are measured across the magnetic field through one of the non-cusp region and at the mid $(r,\theta)$ plane of device (the plane from z = 65 cm from the filament) as shown in figure 4. The argon gas background pressure is $2.0 \times 10^{-4}$ mBar throughout this experiment until and unless specified.

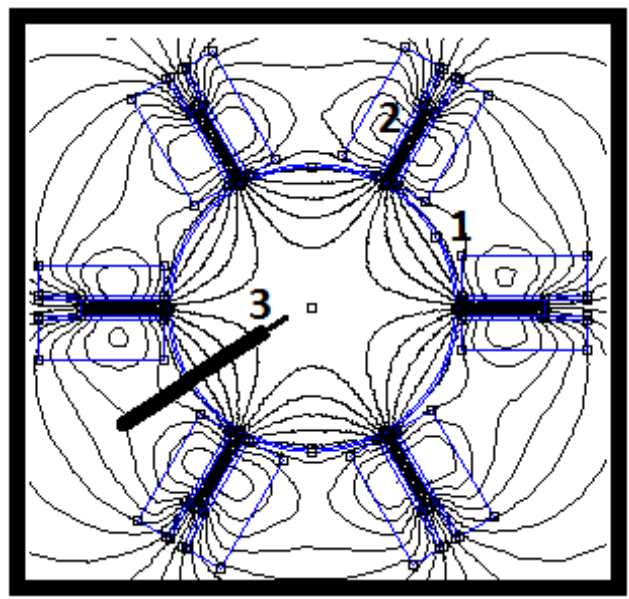

Figure 4. Schematic diagram of z = 65 cm (from the filament), $(r, \theta)$ plane. 1: Chamber cross section, 2: Magnet, 3: Langmuir probe across the magnetic and along non-cusp region field (exactly in between two consecutive magnets).

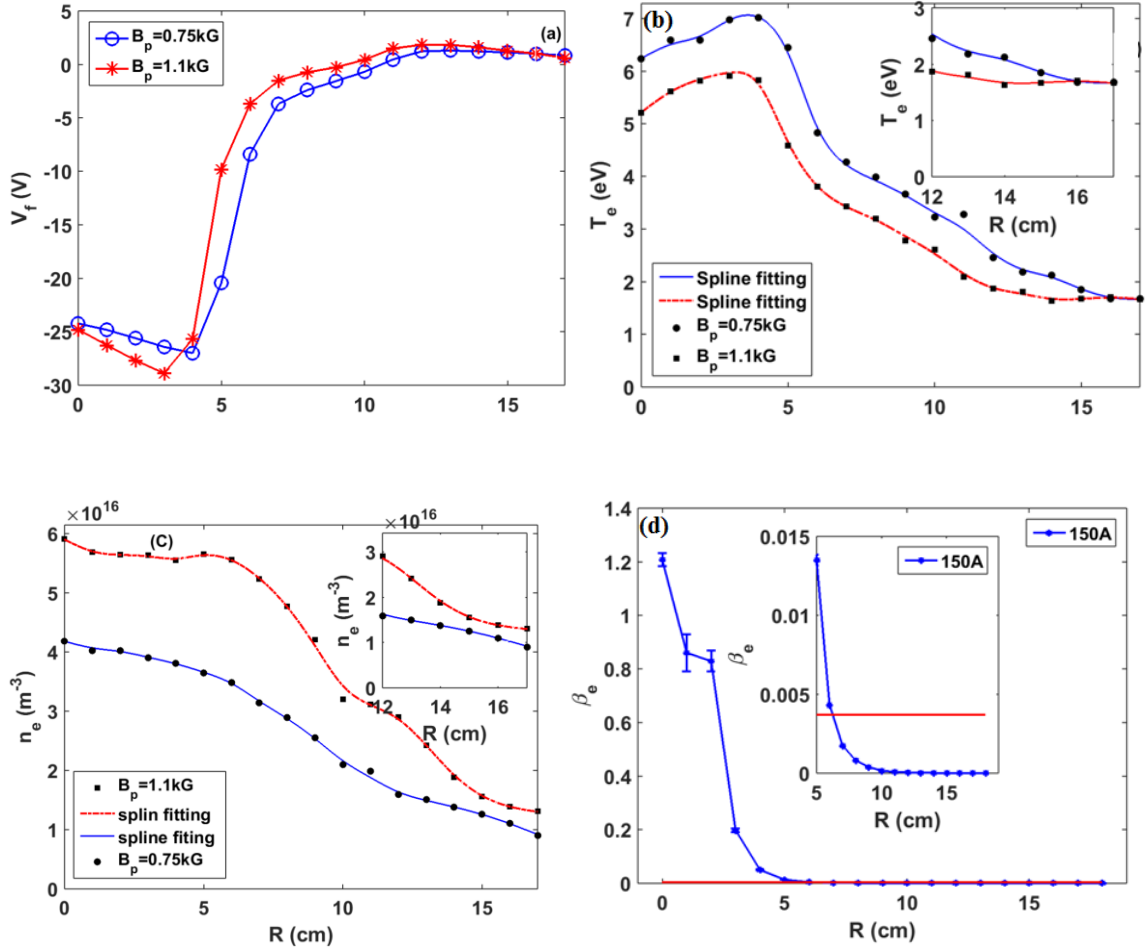

Figure 5. Radial variation of mean plasma parameters, (a) floating potential ($V_f$), (b) electron temperature ($T_e$), and (c) plasma density ($n_e$) at two different pole magnetic fields $B_p$ = 0.75 kG and $B_p$ = 1.1 kG when magnets are energies with magnet currents 100 A and 150 A respectively. (d) Radial variation of plasma beta ($\beta_e$) along non-cusp region when magnets are energized with current ($I_{mag}$) is 150 A and Argon gas pressure $2 \times 10^{-4}$ mBar. The line shows the value of plasma beta line $y = \sqrt{m_e/m_i}$ for reference.

The confined plasma in cusp magnetic field is diffused across the magnetic field lines and stream out along the magnetic field lines because of these diffusion the gradient in equilibrium profile of plasma parameters are formed. The radial variation of mean plasma density, electron temperature, and floating potential are measured along the non-cusp region using simple conventional Langmuir probe at two different pole magnetic fields $B_p$ = 0.75 kG and $B_p$ = 1.1 kG case as shown in figure 5. Since the cusp magnetic field confined the primary electrons in confined region thus the floating potential ($V_f$) is highly negative in the bulk plasma (up to 4cm) as shown in figure 5 (a). The confined primary electrons are scavenged across the magnetic field hence the floating potential decreased with increasing radial distance as shown in figure 5 (a). The diffuse plasma across the magnetic field (from R = 7 to 17 cm) has very low floating potential and it satisfied the condition $V_p = V_f + 5.4\ T_e$. Thus plasma in this region has very low population of primary electrons and has Maxwellian nature. The figure 5 (b) and (c) show the radial variation of electron temperature ($T_e$) and plasma density ($n$) along the non cusp region. The central region of plasma in MPD device is hot (electron temperature is high), uniformly dense and enrich of primary electrons up to 5cm. The confined bulk plasma is diffused along as well as across the magnetic field and because of diffusion gradient in mean plasma density and electron temperature is formed shown in figure 5 (b) and (c).

In figure 5, we have used the smoothing Spline fitting on radial profile of density ($n$) and electron temperature ($T_e$) and estimated the density scale length ($L_n$) and temperature scale length ($L_{Te}$). Here the scale length is defined by the equation; $L_x = |x/(dx/dR)|$, and suffix $x$ is $n$ and $T_e$. Considering these plasma parameters scale lengths we have divided the plasma of the device in three regions. First region, from radial distance R = 0 to 6 cm the plasma density and electron temperature is uniform and plasma is quiescent, in second region, from 6 to 10 cm, $L_n > L_{Te}$ ($L_n = 50$ cm, and $L_{Te} = 6$ cm at R = 7 cm for pole magnetic field $B_p = 0.75$ kG case), the temperature gradient is dominated compared to density gradient and in third region from 12 to 17cm $L_{Te} > L_n$ the density gradient is dominated as compared to temperature gradient. In third region at R = 14 cm $L_n$ and $L_{Te}$ are 12 cm, 50 cm for pole magnetic field $B_p = 0.75$ kG respectively and $L_n$ and $L_{Te}$ are 4.27 cm, 60 cm for pole magnetic field $B_p = 1.1$ kG respectively. The inequality in scale length, $L_{Te} > L_n$ supports the drift wave instability in third region. We limited our study in third region. Figure 7 (d) shows the radial variation of plasma beta ($\beta_e$) along the non-cusp region. In the third region it has been observed that $\beta_e$ is less than $\sqrt{m_e/m_i}$, thus the instability in the third region has electrostatic in nature.

b) **Spectral analysis**

The important characteristic feature such as fluctuation in plasma parameter, power spectra of fluctuation, cross correlation function, wave number-frequency spectrum are necessary to identify the instability. In this regard we have measured fluctuation in plasma density, cross correlation between two physical quantities and wave number.

1. **Fluctuation and frequency spectra analysis**

We now describe the experimental results on fluctuation. The figure 6 (a) shows time profile of plasma density fluctuations at two radial location and for different pole cusp magnetic fields $B_p = 0.75$ kG and $B_p = 1.1$ kG case along the non-cusp region. The density fluctuation is measured from the ion saturation current fluctuation. We used simple Langmuir probe to measured ion saturation current fluctuation across the 10 kΩ resistance with sampling rate 1MSa/sec and record length 1M point using KEYSIGHT DSOX2024A (200 MHz band width and 2 GS s$^{-1}$ sampling rate) CRO. From figure 6 (a), it is observed that at radial distance R = 0 cm, the fluctuation in density is suppressed because the plasma in this region has uniform plasma parameter also confined plasma in this region by a good curvature of magnetic field, but at radial distance R = 14 cm along non-cusp fluctuation in density is present for pole cusp magnetic fields $B_p = 0.75$ kG case because of radial gradient in density, we have also notice that fluctuation in density is increased more at same radial location for pole cusp magnetic fields $B_p = 1.1$ kG case, because the gradient in density is increased .

Figure 6 (b) shows the radial variation of normalised plasma density fluctuation level at two different pole cusp magnetic fields $B_p = 0.75$ kG, and $B_p = 1.1$ kG case when magnets are energies with magnet currents 100 A, and 150 A respectively. It is clearly observed from figure that radial variation of normalized density fluctuation level follows the mean plasma density profile for both cases. The maximum normalized density fluctuation level are 6% and 11 % for pole cusp magnetic fields $B_p = 0.75$ kG, and $B_p = 1.1$ kG case respectively. The normalized density fluctuation level is high for pole cusp magnetic fields $B_p = 1.1$ kG because of the gradient in density is high for $B_p = 1.1$ kG case. The normalized density fluctuations follow the density profile as well as increased in density fluctuation with density gradient [33]. We have also noticed that in second region of device (R = 5 cm to 11cm) there is a gradient in temperature as well as density but fluctuation in density is suppressed compared to third region. The reason behind suppression in density fluctuation is plasma confined in magnetic fields have good curvature in that regime. Plasma confined in good curvature of magnetic field suppresses the instability because magnetic field gradient drift and curvature drift towards the plasma [34], but in edge (third) regime density gradient and bad curvature of magnetic field enhance the fluctuation in density.

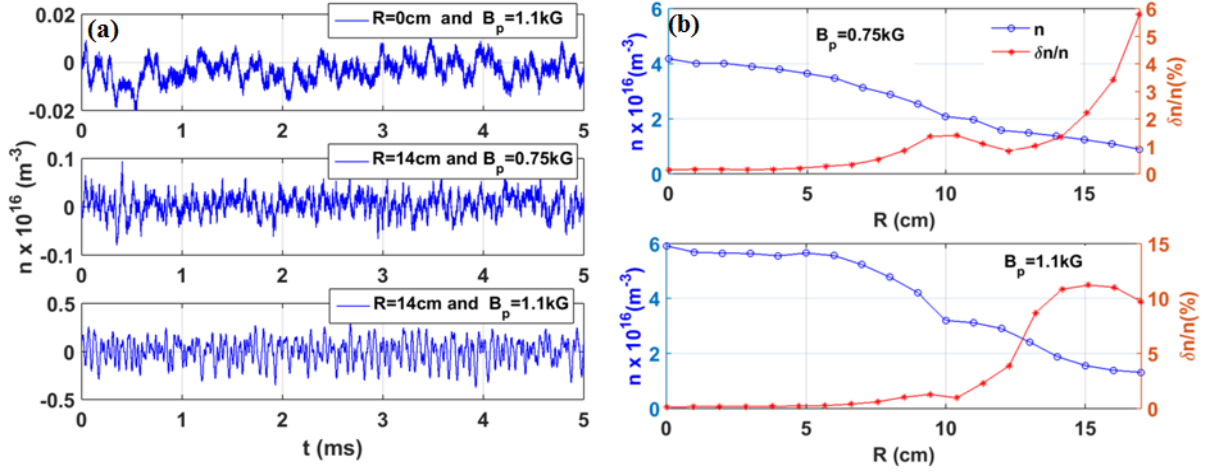

Figure 6. (a) Time profile of density fluctuation, at R = 0 cm and at R = 14 cm for two different pole cusp magnetic fields $B_p$ = 0.75 kG and $B_p$ = 1.1 kG when magnets are energies with magnet currents 100 A, and 150 A respectively, (b) radial profile of normalized density fluctuation level for two different pole cusp magnetic fields $B_p$ = 0.75 kG and $B_p$ = 1.1 kG when magnets are energies with magnet currents 100 A, and 150 A respectively.

The important characteristic features such as power spectra and cross correlation function is necessary to identify the instability. Figure 7 (a) shows the auto power spectrum of density fluctuation. It is observed that power spectra has mean frequency peak at 3 kHz for pole cusp magnetic fields $B_p$ = 0.75 kG case and the power spectra become broad band with significant power and central frequency at 15 kHz for pole cusp magnetic fields $B_p$ = 1.1 kG case. The frequency is shifted from 3 kHz to 15 kHz because of increasing in density gradient. The power spectra also follows a power law $1/f^{5.5}$ for $f \leq 20$ kH to 70 kHz for density fluctuations shown in figure 7 (b) [35].

The cross correlation function between the normalized values of density fluctuation ($\delta n/n$) with floating potential ($eV_f / k_B T_e$) is found to be strongly correlated for both cases as shown in figure 8 for both cases. The correlation coefficients obtained is $c(\tau) \sim 0.8$. Also the phase angle between density and floating potential fluctuation is nearly $45^0$ [36-37]. These measurements are carried out using two simple Langmuir probes of dimension 5 mm length and 0.5 mm diameter tungsten tip and distance between two Langmuir probe is 5 mm. There may be a slight spatial de-correlation as the probes used are not located on the same radial location. The density fluctuation measured from ion saturation current fluctuation across 10 kΩ resistance and floating potential fluctuation measured across 5 MΩ resistance with sampling rate 1 MS s$^{-1}$ and record length 1M point using same **CRO**.

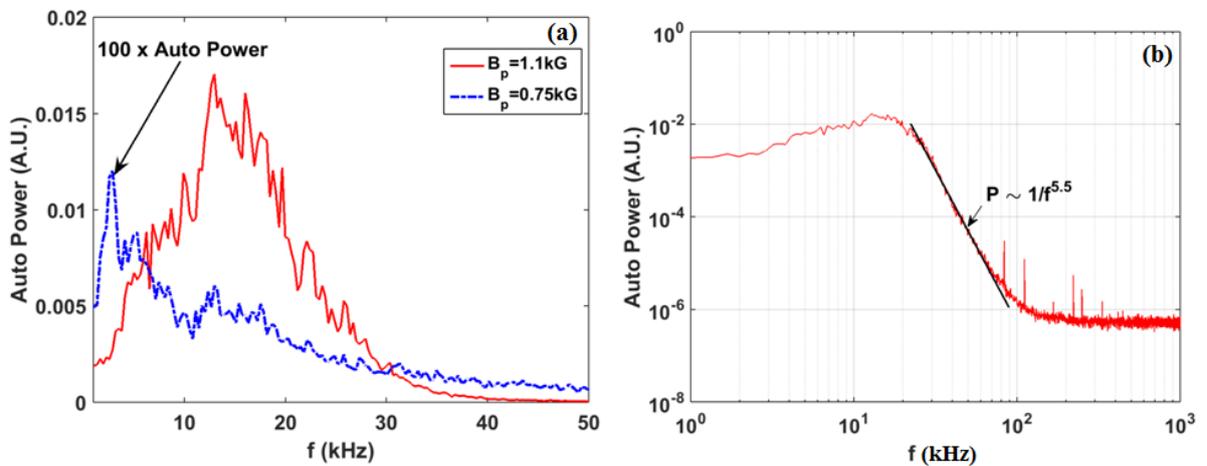

Figure 7. (a) Auto-power spectrum of density at R = 14 cm and two different pole cusp magnetic fields $B_p$ = 0.75 kG and $B_p$ = 1.1 kG when magnets are energies with magnet currents 100 A, and 150 A respectively. (b) Log-log scale variation for Auto-power spectrum of plasma density fluctuation.

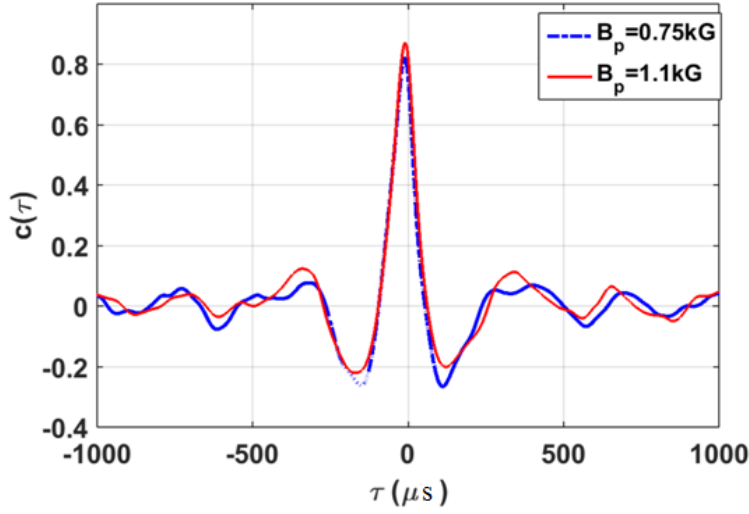

**Figure 8:** (a) The cross correlation function c (τ) between normalized time series of density and floating potential fluctuation at R = 14 cm and two different pole cusp magnetic fields $B_p$ = 0.75 kG, and $B_p$ = 1.1 kG when magnets are energies with magnet currents 100 A, and 150 A respectively.

2.  **Wavenumber-frequency spectra analysis**

Figure 9 represents a schematic diagram of drift wave velocity along z-axis (a) in (r, θ) plane and (b) in (r, z) plane. From this figure we understand that the direction of the drift wave changes in the alternate non-cusp region due to the change in the magnetic field direction. In figure 10 the contour plots of wave number-frequency spectrum $S(k_z, \omega)$ clearly captures this phenomenon. Figures have been plotted using density fluctuations, measured from two simple Langmuir probes separated by 1cm along the z-axis at radial location R = 14 cm for two different pole cusp magnetic fields $B_p$ = 0.75 kG [10.(a) & 10.(b) for alternate non-cusp region] & $B_p$ = 1.1 kG [10.(c) & 10.(d) for alternate non-cusp region] with sampling rate 2 MS s$^{-1}$, and record length 5 M point using a same CRO. Then the drift velocity ($V_d$) is calculated by noting the associated wave number ($k_z$) at the peak frequency (ω) from the wave number-frequency spectrum $S(k_z, \omega)$ and compared with the calculated drift velocity ($V_d$) from density scale length formula $V_d = k_B T_e / eBL_n$ [24, 33, 37].

For $B_p$ = 0.75 kG, the spectrum $S(k_z, \omega)$ shows peak at frequency 3 kHz as shown in figure 10 (a) and its associated wave number ($k_z$) and wavelength (λ) of this mode are 0.11565 cm$^{-1}$ and 54 cm respectively. The drift velocity obtained from $S(k_z, \omega)$ spectrum becomes 1.15 x 10$^5$ cm/s which matches the theoretically calculated value 1.0 x 10$^5$ cm/s , obtained from the density scale length formula at radial location R = 14 cm and thus vindicating our experimental results. From figure 10 (a) and (b) it has been observed that at the alternate non-cusp region, the direction of the drift velocity is a change its sign which leads to negative wave number. Similarly, for $B_p$ = 1.1 kG the spectrum $S(k_z, \omega)$ broadens at central frequency 15 kHz, as shown in figure 10 (c), and the associated wave number ($k_z$) & wavelength (λ) of this mode become 0.3665 cm$^{-1}$ & 17 cm respectively. The drift velocity is obtained from the slope of line fitted at broadband frequency of $S(k_z, \omega)$ [as shown in figure 10 (c)]. In this case the drift velocity from $S(k_z, \omega)$ spectrum and calculated from the density scale length at radial location R = 14 cm are 2.9 x 10$^5$ cm/s and 2.55 x 10$^5$ cm/s respectively. Figure 10.(d) shows the wave-number frequency spectra $S(k_z, \omega)$ opposite non-cusp region. The drift velocity observed at peak frequency from spectrum $S(k_z, \omega)$ is nearly matched with calculated from density scale length for m=1 drift wave mode and the wavelength of mode (λ) is also very large compared to ion gyro radius ($\rho_i$) for both cases.

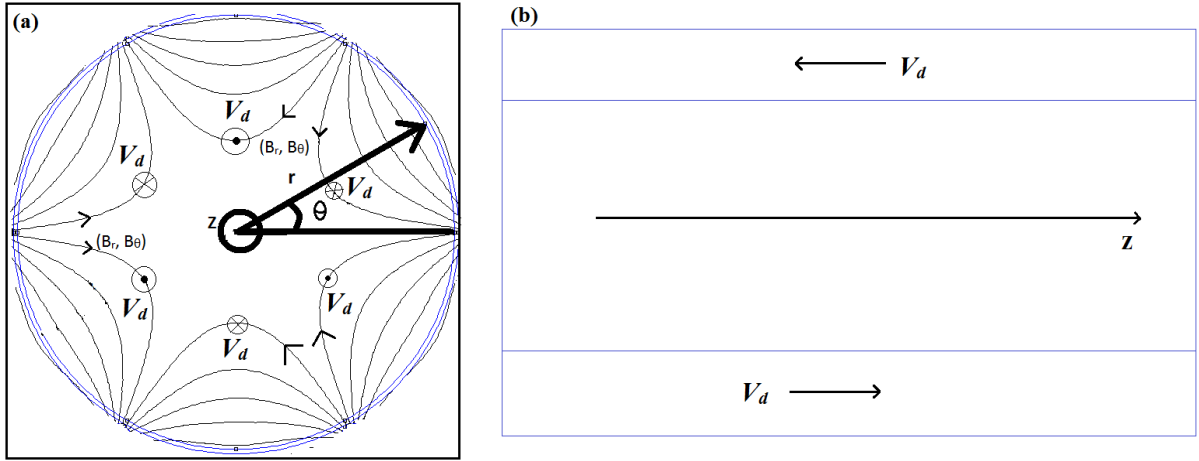

**Figure 9:** Schematic diagram of drift velocity ($V_d$) along the z-axis (a) in ($r$, $\theta$) plane and (b) in ($r$, $z$) plane of the MPD device.

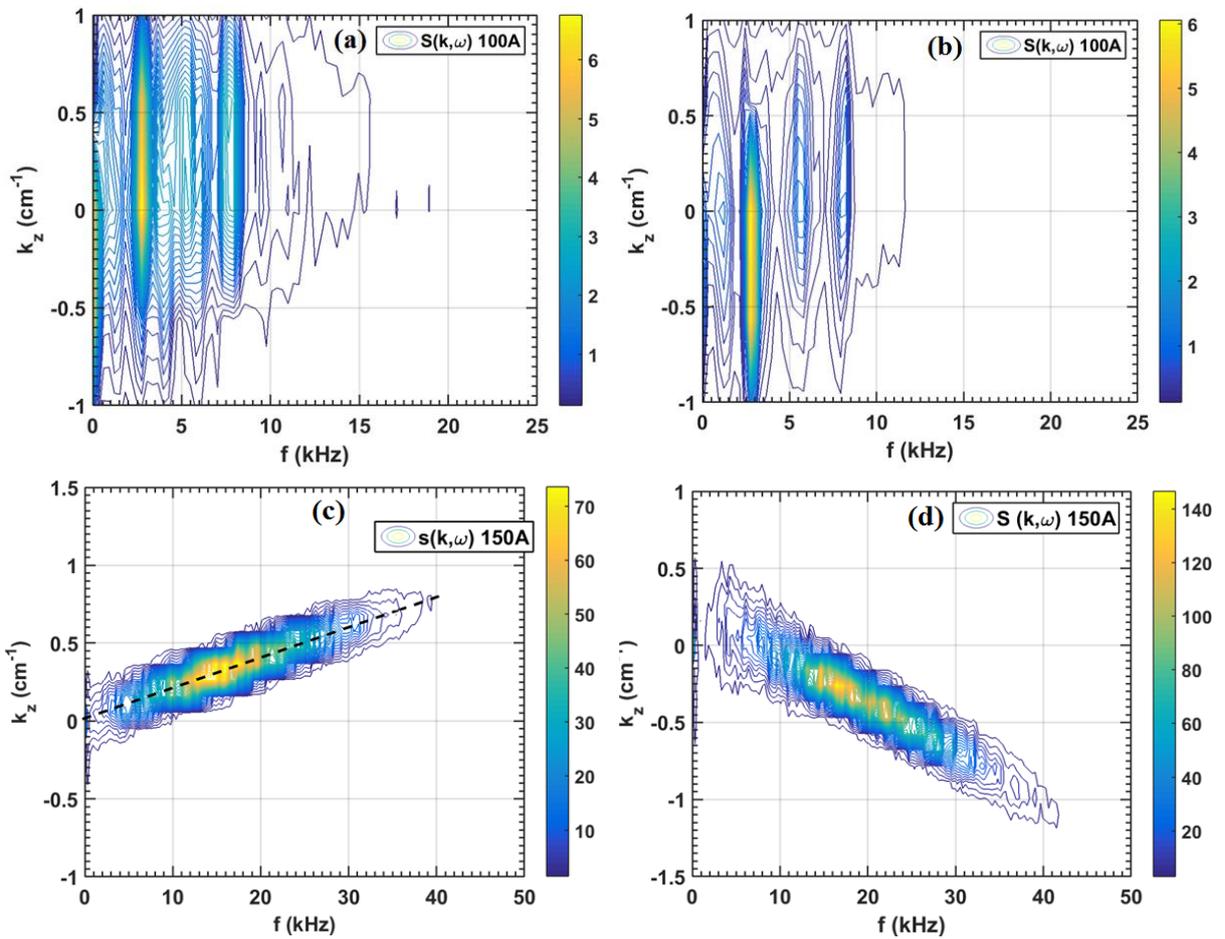

Figure 10: The countour plot of joint wave number and frequency specturm at R = 14 cm and two different pole cusp magnetic fields $B_p$=0.75 kG, and $B_p$ = 1.1 kG when magnets are energies with magnet currents 100 A, and 150 A respectively.

## IV. Summary and Conclusion

In this paper we have reported the experimental observations of density gradient driven drift wave turbulence in low-β plasma confined by an inhomogeneous cusp magnetic field. The equilibrium profile of plasma can be controlled by the magnetic field of the device. Out of this equilibrium profiles we have studied the drift wave instability for two extreme equilibrium profiles. The generated instability has been identified by explicitly studying the following properties:

- Comparison of scale lengths of plasma parameters, as for example inequality in scale length, $L_{Te} > L_n$ support the condition for drift wave instability.
- In time evolution of density profile, fluctuations have been observed when the density gradient present, whereas these fluctuations are suppressed in uniform plasma confined region. This radial variation of normalized density fluctuation level follows the density profile which excites drift wave instability.
- The density fluctuation frequency power spectra has power index 5.5 [35].
- The strongly correlated cross correlation function between normalized floating poetical fluctuation and density fluctuation hold the relation $(\delta n/n) \approx (eV_f/k_B T_e)$ and nearly π/4 phase angle between them both are prove that the density fluctuation lead floating potential fluctuation for density gradient driven drift wave turbulence.
- Lastly the cross field drift velocity due to fluctuation in plasma parameters have been measured from the wave number- frequency $S(k_z, \omega)$ spectrums and are compared with the theoretical values, obtained from density scale length formula. Also the wavelength of mode ($\lambda$) is very large compared to ion gyro radius ($\rho_i$) which leads to "anomalous diffusion" across the inhomogeneous magnetic field.

Thus all above diagnostics confirm the instability in non-cusp region is drift wave instability due to density gradient along the radius and across the magnetic field ($B_\theta$). Thus the particles drift due to density fluctuation along the axial direction (z-direction). Moreover it has been first time observed that, the direction of the drift wave velocity changes its sign in alternate non–cusp region.

## Acknowledgment:

The authors would like to express sincere gratitude to Dr. Rajwinder Kaur Sengupta for giving valuable suggestions for improving the manuscript as well as reviewing the manuscript. The author A. D. P also acknowledges Mr. Lavkesh Lachhvani for helping analysis of Wave number-frequency spectrum analysis. The author A. D. P also acknowledges Mr. Arghya Mukherjee for correcting the manuscript.